\documentclass[conference]{IEEEtran}
\IEEEoverridecommandlockouts
\usepackage{cite}
\usepackage{amsmath,amssymb,amsfonts}
\usepackage{algorithmic}
\usepackage{graphicx}
\usepackage{textcomp}
\usepackage{xcolor}
\def\BibTeX{{\rm B\kern-.05em{\sc i\kern-.025em b}\kern-.08em
    T\kern-.1667em\lower.7ex\hbox{E}\kern-.125emX}}
\usepackage{url}
\usepackage{booktabs} 
\usepackage{algorithmic}
\usepackage{comment}
\usepackage{colortbl}
\usepackage[ruled]{algorithm2e} 

\usepackage{subcaption}
\usepackage{bm}
\usepackage{multirow}
\usepackage{caption}
\usepackage{fancyhdr}
\def\red{\textcolor{red}}

\makeatletter
\def\endthebibliography{%
  \def\@noitemerr{\@latex@warning{Empty `thebibliography' environment}}%
  \endlist
}
\makeatother

\begin{document}

\title{Know Your Mind: Adaptive Cognitive Activity Recognition with Reinforced CNN}

\author{\IEEEauthorblockN{Xiang Zhang$^*$, Lina Yao$^*$, Xianzhi Wang$^\S$, Wenjie Zhang$^*$, Shuai Zhang$^*$, Yunhao Liu$^\P$}
\IEEEauthorblockA{$^*$University of New South Wales, Sydney, Australia \\ 
$^\S$ University of Technology Sydney, Sydney, Australia\\
$^\P$ Michigan State University, East Lansing, USA\\
\{xiang.zhang3, shuai.zhang\}@student.unsw.edu.au, \{lina.yao, wenjie.zhang\}@unsw.edu.au \\
xianzhi.wang@uts.edu.au, yunhao@cse.msu.edu}
}

\maketitle

\begin{abstract}
Electroencephalography (EEG) signals reflect and measure activities in certain brain areas. Its zero clinical risk and easy-to-use features make it a good choice of providing insights into the cognitive process. However, effective analysis of time-varying EEG signals remains challenging. First, EEG signal processing and feature engineering are time-consuming and highly rely on expert knowledge, and most existing studies focus on domain-specific classification algorithms, which may not apply to other domains. Second, EEG signals usually have low signal-to-noise ratios and are more chaotic than other sensor signals. 
In this regard, we propose a generic EEG-based cognitive activity recognition framework that can adaptively support a wide range of cognitive applications to address the above issues. The framework uses a reinforced selective attention model to choose the characteristic information among raw EEG signals automatically. It employs a convolutional mapping operation to dynamically transform the selected information into a feature space to uncover the implicit spatial dependency of EEG sample distribution. We demonstrate the effectiveness of the framework under three representative scenarios: intention recognition with motor imagery EEG, person identification, and neurological diagnosis, and further evaluate it on three widely used public datasets.
The experimental results show our framework outperforms multiple state-of-the-art baselines and achieves competitive accuracy 
on all the datasets while achieving low latency and high resilience in handling complex EEG signals across various domains. The results confirm the suitability of the proposed generic approach for a range of problems in the realm of Brain-Computer Interface applications. 
\end{abstract}


\section{Introduction} 
\label{sec:introduction}
Electroencephalography (EEG) is an electrophysiological monitoring indicator to analyze brain states and activities by measuring the voltage fluctuations of ionic current within the neurons of brains \cite{zhang2019survey}. In practice, EEG signals can be collected by portable and off-the-shelf equipment in a non-invasive and non-stationary way \cite{adeli2007wavelet}.
EEG signal classification algorithms have been studied for a range of real-world applications \cite{zhang2017eeg}. 
The accuracy and robustness of EEG classification model have promising meanings to identify cognitive activities in the realms of movement intention recognition, person identification, and neurological diagnosis. 
Cognitive activity recognition systems \cite{vallabhaneni2005brain} provide a  bridge between the inside cognitive world and the outside physical world. They are recently used in assisted living \cite{zhang2017converting}, smart homes \cite{zhang2017intent}, and entertainment industry \cite{russoniello2009effectiveness}; EEG-based person identification technique empowers the security systems deployed in bank or customs \cite{schetinin2017feature,zhang2018mindid}; EEG signal-based neurological diagnosis can be used to detect the organic brain injury and abnormal synchronous neuronal activity such as epileptic seizure \cite{veeriah2015deep,acar2007seizure}. 

\begin{figure}[t]
    \centering
    \begin{subfigure}[t]{0.1\textwidth}
        \centering
        \includegraphics[width=\textwidth]{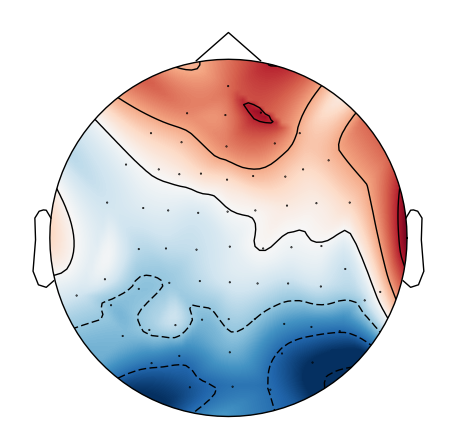}
        \caption{T-2}
    \end{subfigure}%
    \begin{subfigure}[t]{0.1\textwidth}
      \centering
      \includegraphics[width=\textwidth]{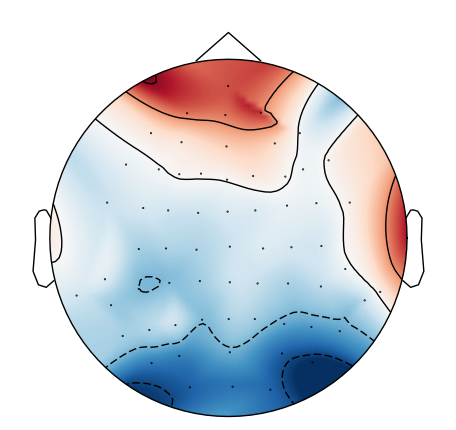}
      \caption{T-1}
    \end{subfigure}%
    \begin{subfigure}[t]{0.1\textwidth}
        \centering
        \includegraphics[width=\textwidth]{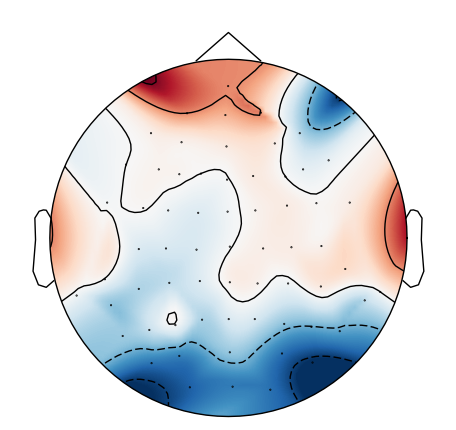}
        \caption{T}
    \end{subfigure}%
    \begin{subfigure}[t]{0.1\textwidth}
        \centering
        \includegraphics[width=\textwidth]{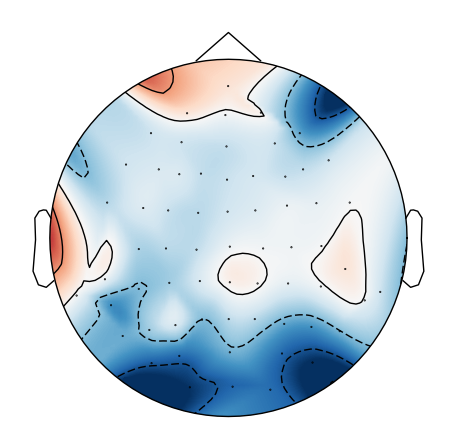}
        \caption{T+1}
    \end{subfigure}%
    \begin{subfigure}[t]{0.1\textwidth}
        \centering
        \includegraphics[width=\textwidth]{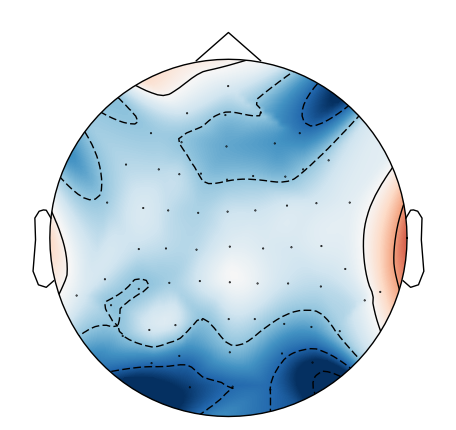}
        \caption{T+2}
    \end{subfigure}%
    \caption{EEG topography with continuous samples. The interval among samples is 0.00625 second.}
    \label{fig:topography}
    \vspace{-3mm}
\end{figure}

\begin{table*}[t]
\centering
\caption{Time domain and correlation coefficient analysis. \textit{n-points} denotes the values are measured by the samples with \textit{n} sampling points. We compare EEG signals with other sensing data (such as wearable sensor data and smartphone data) over five different scales and the results constantly show that EEG signals have the highest instability.}
\label{tab:eeg_analysis}
\begin{tabular}{llllllllllllll}
\hline
\multirow{5}{*}{\textbf{\begin{tabular}[c]{@{}l@{}}Time \\ Domain\end{tabular}}} & \multirow{2}{*}{\textbf{Signals}} & \multicolumn{2}{c}{\textbf{5-points}} & \multicolumn{2}{c}{\textbf{50-points}} & \multicolumn{2}{c}{\textbf{100-points}} & \multicolumn{2}{c}{\textbf{500-points}} & \multicolumn{2}{c}{\textbf{1000-points}} & \multicolumn{2}{c}{\textbf{Average}} \\
 &  & \textbf{STD} & \textbf{Range} & \textbf{STD} & \textbf{Range} & \textbf{STD} & \textbf{Range} & \textbf{STD} & \textbf{Range} & \textbf{STD} & \textbf{Range} & \textbf{STD} & \textbf{Range} \\ \cline{2-14}
 & \textbf{Phone} & 0.0025 & 0.0061 & 0.0179 & 0.0494 & 0.0166 & 0.0612 & 0.0253 & 0.1177 & 0.0259 & 0.1281 & 0.0882 & 0.3625 \\
 & \textbf{Wearable} & 0.0012 & 0.0029 & 0.0107 & 0.0369 & 0.0147 & 0.0519 & 0.0197 & 0.1041 & 0.016 & 0.1058 & 0.0623 & 0.3016 \\
 & \textbf{EEG} & 0.0087 & 0.0218 & 0.0199 & 0.0824 & 0.0245 & 0.1195 & 0.0299 & 0.1619 & 0.0308 & 0.1802 & \textbf{0.1138} & \textbf{0.5658} \\ \hline \hline
\multirow{5}{*}{\textbf{\begin{tabular}[c]{@{}l@{}}Correlation\\ Coefficient\end{tabular}}} & \multirow{2}{*}{\textbf{Signals}} & \multicolumn{2}{c}{\textbf{5-points}} & \multicolumn{2}{c}{\textbf{50-points}} & \multicolumn{2}{c}{\textbf{100-points}} & \multicolumn{2}{c}{\textbf{500-points}} & \multicolumn{2}{c}{\textbf{1000-points}} & \multicolumn{2}{c}{\textbf{Average}} \\
 &  & \textbf{STD} & \textbf{Range} & \textbf{STD} & \textbf{Range} & \textbf{STD} & \textbf{Range} & \textbf{STD} & \textbf{Range} & \textbf{STD} & \textbf{Range} & \textbf{STD} & \textbf{Range} \\ \cline{2-14}
 & \textbf{Phone} & 0.0015 & 0.0038 & 0.0243 & 0.0832 & 0.0248 & 0.0964 & 0.0244 & 0.104 & 0.0247 & 0.104 & 0.0997 & 0.3914 \\
 & \textbf{Wearable} & 0.01 & 0.0252 & 0.0155 & 0.0702 & 0.0147 & 0.0866 & 0.0469 & 0.2299 & 0.0729 & 0.3905 & 0.16 & 0.8024 \\
 & \textbf{EEG} & 0.0392 & 0.0991 & 0.1077 & 0.4096 & 0.0955 & 0.4849 & 0.1319 & 0.7626 & 0.1533 & 0.99 & \textbf{0.5276} & \textbf{2.7462} \\ \hline
\end{tabular}
\vspace{-3mm}
\end{table*}

The classification of cognitive activity faces several challenges. First, the EEG data preprocessing and feature extraction methods (e.g., filtering, Discrete Wavelet Transformation, and feature selection) which are employed by most existing EEG classification studies \cite{zhang2017eeg,russoniello2009effectiveness} 
are time-consuming and highly depend on expertise. Meanwhile, the hand-crafted features require extensive experiments to generalize well to diverse settings such as filtering bands and wavelet orders. Therefore, an effective method which can directly work on raw EEG data is necessary. 

Second, most current EEG classification methods are designed based on domain-specific knowledge and thus may become ineffective or even fail in different scenarios \cite{adeli2007wavelet}. For example, the approach customized for EEG-based neurological diagnosis may not work well on intention recognition. Therefore, a general EEG signal classification method is expected to be both efficient and robust across various domains for better usability and suitability.

Third, EEG signals have a low signal-to-noise ratio and more chaotic than other sensor signals such as wearable sensors.
Thus, the segment-based classification which is widely used in sensing signal classification may not fit cognitive activity recognition. 
A segment contains some continuous EEG samples clipped by the sliding window method \cite{fraschini2015eeg} while a single EEG sample (also called EEG instance) is collected at a specific time point.
In particular, segment-based classification has two drawbacks compared with sample-based classification: 1) in a segment with many samples, the sample diversity may offset by other inverse changed samples as EEG signals vary rapidly (Section~\ref{sec:eeg_characteristic_analysis}). 
2) segment-based classification requires more training data and a longer data-collecting time. For example, suppose each segment has ten samples without overlapping; for the same training batch size, segment-based classification requires ten times of the data size and the data-collecting time than sample-based classification. As a result, segment-based classification cannot exploit the immediate intention of changing and thus achieves low precision in practical deployment. To this end, sample-based classification is more attractive. 


To address the aforementioned issues, 
first, we propose a novel framework which can automatically learn distinctive features from raw EEG signals by developing a deep convolutional mapping component. Additionally, to grasp the characteristic information from different EEG application circumstance adaptively, we design a reinforced selective attention component that combines the benefits of attention mechanism \cite{zhangmulti} and deep reinforcement learning. Moreover, we overstep the challenge of chaotic information by working on EEG samples instead of segments. The single EEG sample only contains spatial information without spatial clue\footnote{We do not deny the usefulness of temporal information, but this paper emphasizes on spatial information, which may easier to be captured.}. 
The main contributions of this work are highlighted as follows:
\begin{itemize}
  \item We propose a general framework for automatic cognitive activity recognition to facilitate a scope of diverse cognitive application domains including intention recognition, person identification, and neurological diagnosis. 
  \item We design the reinforced selective attention model, by combining the deep reinforcement learning and attention mechanism, to automatically extract the robust and distinct deep features. Specially, we design a non-linear reward function to encourage the model to select the best attention area that leads to the highest classification accuracy. Besides, we customize the states and actions based on our cognitive activity recognition environment.
  \item We develop a convolutional mapping method to explore the distinguishable spatial dependency and feed it to the classifier for classification, among selected EEG signals. 
  \item We demonstrate the effectiveness of the proposed framework using four real-world datasets concerning three representatives and challenging cognitive applications. The experiment results demonstrate that the proposed framework outperforms the state-of-the-art and strong baselines by consistently achieving the accuracy of more than 96\% and low latency. 
\end{itemize}

Note that all the necessary reusable codes and datasets have been open-sourced for reproduction, please refer to this link\footnote{\url{https://github.com/xiangzhang1015/know_your_mind}}.


\section{Analysis of EEG Signals} 
\label{sec:eeg_characteristic_analysis}
In this section, we demonstrate EEG signals' unique characteristics (e.g., rapid-varying and chaotic) and that single samples are more suitable than segments for classification.
By comparing EEG signals with two typical sensor signals collected by smartphone (accelerometers in Samsung Galaxy S2) and wearable sensors (Colibri wireless IMU).
The participants are walking during the data collection session.


The brain activity is very complex and rapid varying, but EEG signals can only capture a few information through the discrete sampling of biological current. Figure~\ref{fig:topography} demonstrates the characteristics of rapidly varying and complex of EEG signals and provides the EEG topography of consecutive 5 samples. The sampling rate is 160 Hz while the sampling interval is 0.00625 second. It can be observed that the topography changes dramatically within such a tiny time interval. 




Furthermore, to illustrate the chaotic of EEG signals, we compare EEG with smartphone and wearable sensors in two aspects: the time domain and the inter-samples correlations.


In the time domain, we evaluate the STD and range of sensor signals on five levels of sample length: 5, 50, 100, 500, 1000 continuous samples. 
The evaluations on the above five scales are expected to show the tendency that how the EEG characteristic varies with the sampling period. 

The inter-sample correlation coefficient calculates the average cosine correlations between the specific sample and its neighbor samples (5, 50, 100, 500, and 1000 samples).
A low correlation coefficient represents EEG signals dramatically and rapidly varying all the time. 

As a result, Table~\ref{tab:eeg_analysis} present the STD and range values in the time domain and correlation coefficient. We observe that EEG signals have the highest STD and range over all the five sample window scales both on time domain and correlation coefficient, compared with wearable sensor data and smartphone signals. This demonstrates that the EEG sample has more unstable correlations with neighbors and the instability is very high even in the nearest five samples. More specifically, EEG signals are very chaotic and rapidly changing at each single sampling point. 

\begin{figure*}[t]
\centering
  \includegraphics[width=0.9\linewidth]{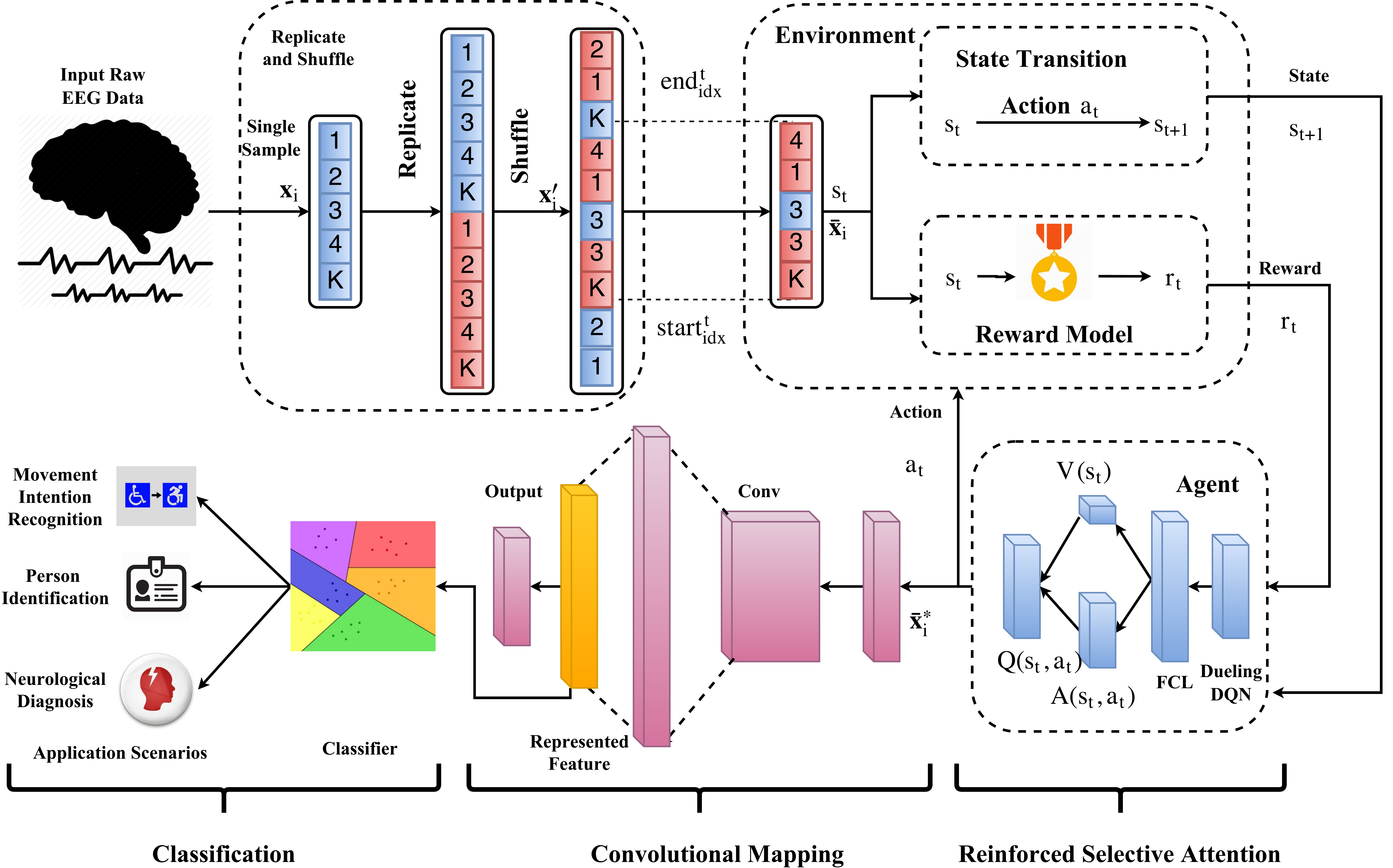}
  \caption{Flowchart of the proposed approach. The input raw EEG single sample $\mathbf{x}_i$ (K denotes the $K$th element) is replicated and shuffled to provide more latent spatial combinations of feature dimensions. Then, an attention zone $\mathbf{\bar{x}}_i$, which is a fragment in $\mathbf{x}'_i$, with the state $s_t = \{start^t_{idx}, end^t_{idx}\}$ is selected. The selected attention zone is input to the state transition and the reward model. In each step $t$, one action is selected by the state transition to update $s_t$ based on the agent's feedback. The reward model evaluates the quality of the attention zone by the reward score $r_t$. The dueling DQN is employed to discover the best attention zone $\mathbf{\bar{x}}^*_i$ which will be fed into the convolutional mapping procedure to extract the spatial dependency representation. The represented features will be used for the classification. $FCL$ denotes a fully connected layer. The reward model is the combination of the convolutional mapping and the classifier.
   }
  \label{fig:workflow}
\end{figure*}

\section{Proposed Method}
\label{sec:methodology}
Based on the above analysis, we propose reinforced attentive convolutional neural networks (CNNs) to classify raw EEG signals accurately and efficiently directly.
The overall workflow is shown in Figure~\ref{fig:workflow}. 

\begin{figure}[]
    \centering
    \includegraphics[width=0.45\textwidth]{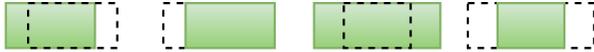}
 \caption{Four actions in the state transition: left shifting, right shifting, extend, and condense. 
  The dashed line indicates the position of the attention zone before the action while the solid line
  indicates after the action.}
\label{fig:actions}
\end{figure}

\subsection{Replicate and Shuffle} 
\label{sub:repeat_and_shuffle}
To provide as much as possible information, we design an approach to exploit the spatial relationships among EEG signals. The signals belonging to different brain activities are supposed to have different spatial dependent relationships.
We replicate and shuffle the input EEG signals on dimension-wise. Within this method, all the possible dimension arrangements have the equiprobable appearance. 

Suppose the input raw EEG data are denoted by $\mathbf{X}=\{(\mathbf{x}_i, y_i), i=1,2,\cdots I\}$, where $(\mathbf{x}_i, y_i)$ denotes a single EEG sample and $I$ denotes the number of samples. In each sample, the feature $\mathbf{x}_i=\{x_{ik},k=1,2,\cdots, K\}, \mathbf{x}_i\in \mathbb{R}^K$ contains $K$ elements corresponding to $K$ EEG channels and $y_i \in \mathbb{R}$ denotes the corresponding label. $x_{ik}$ denotes the $k$-th dimension value in the $i$-th sample.

In real-world collection scenarios, the EEG data are generally concatenated following the distribution of biomedical EEG channels. However, the biomedical dimension order may not present the best spatial dependency.
 The exhausting method is too computationally expensive to exhaust all the possible dimension arrangements. For example, a 64-channel EEG sample has $A^{64}_{64}=1.28\times 10^{89}$ combinations, which is an astronomical figure. 

To provide more potential dimension combinations, we propose a method called Replicate and Shuffle (RS). RS is a two-step mapping method which maps $\mathbf{x}_i$ to a higher dimensional space $\mathbf{x}'_i$ with complete element combinations:
\begin{equation}
\mathbf{x}_i\in \mathbb{R}^K \rightarrow \mathbf{x}'_i \in \mathbb{R}^{K'}, K'>K
\end{equation}

In the first step (Replicate), replicating $\mathbf{x}_i$ for $h = K'/K+1$ times.
 Then, we get a new vector with length as $h*K$ which is not less than $K'$; in the second step (Shuffle), we randomly shuffle the replicated vector in the first step and intercept the first $K'$ elements to generate $\mathbf{x}'_i$. Theoretically, compared with $\mathbf{x}_i$, $\mathbf{x}'_i$ contains more diverse dimension combinations.
Note, this RS operation only be performed once for a specific input dataset in order to provide a stable environment for the following reinforcement learning. 

 \subsection{Reinforced Selective Attention} 
 \label{sub:attention_pattern_learning}
Inspired by the fact that the optimal spatial relationship only depends on a subset of feature dimensions, we introduce an attention zone to focus on a fragment of feature dimensions. Here, the attention zone is optimized by deep reinforcement learning, which has been proved to be stable and well-performed in policy learning.

In particular, we aim to detect the optimal dimension combination, which includes the most distinctive spatial dependency among EEG signals. Since $K'$, the length of $\mathbf{x}'_i$, is too large and computationally expensive, to balance the length and the information content, we introduce the attention mechanism \cite{cavanagh1992attention} since its effectiveness has been demonstrated in recent research areas such as speech recognition \cite{chorowski2015attention}.
We attempt to emphasize the informative fragment in $\mathbf{x}'_i$ and denote the fragment by $\mathbf{\bar{x}}_i$, which is called \textit{attention zone}.
Let $\mathbf{\bar{x}}_i \in \mathbb{R}^{\bar{K}}$ and $\bar{K}$ denote the length of the attention zone which is automatically learned by the proposed algorithm.
 We employ deep reinforcement learning to discover the best attention zone \cite{mnih2015human}. 

As shown in Figure~\ref{fig:workflow}, the detection of the best attention zone includes two key components: the environment (including state transition and reward model) and the agent. Three elements (the state $s$, the action $a$, and the reward $r$) are exchanged in the interaction between the environment and the agent. All of the three elements are customized based on our context in this study. Next, we introduce the design of the crucial components of our deep reinforcement learning structure:

\begin{itemize}
\item The \textbf{state} $\mathcal{S}=\{s_t, t=0,1,\cdots,T\}, s_t \in \mathbb{R}^2$ describes the position of the attention zone, where $t$ denotes the time stamp. 
Since the attention zone is a shifting fragment on 1-D $\mathbf{x}'_i$, we design two parameters to define the state: $s_t = \{start^t_{idx},end^t_{idx}\}$, where $start^t_{idx}$ and $end^t_{idx}$ denote the start index and the end index of the attention zone\footnote{For example, for a random $\mathbf{x}'_i = [3,5,8,9,2,1,6,0]$, the state $\{start^t_{idx}=2, end^t_{idx}=5\}$ is sufficient to determine the attention zone as $[8,9,2,1]$.}, separately.
In the training, $s_0$ is initialized as
\begin{equation}
s_0=[(K'-\bar{K})/2, (K'+\bar{K})/2]
\end{equation}

\item The \textbf{action} $\mathcal{A}=\{a_t,t=0,1,\cdots,T\}\in \mathbb{R}^4$ describes which action the agent could choose to act on the environment. Here at time stamp $t$, the state transition chooses one action to implement following the agent's policy $\pi$: 
\begin{equation}
s_{t+1}=\pi(s_t,a_t)
\end{equation}

In our case, we define four categories of actions (Figure~\ref{fig:actions}) for the attention zone: left shifting, right shifting, extend, and condense.
For each action, the attention zone moves a random distance $d \in[1, d^u]$ where $d^u$ is the upper boundary. 
For left shifting and right shifting actions, the attention zone shifts light-ward or right-ward with the step $d$; for the extend and condense actions, both $start^t_{idx}$ and $end^t_{idx}$ are moving $d$. At last, if the state start index or end index is beyond the boundary, a clip operation is conducted. For example, if $start^t_{idx}=-5$ which is lower than the lower boundary $0$, we clip the start index as $start^t_{idx}=0$.

\item The \textbf{reward} $\mathcal{R}=\{r_t,t=0,1,\cdots,T\}\in \mathbb{R}$ is calculated by the reward model, which will be detailed later. The reward model $\Phi$:
\begin{equation}
r_{t}=\Phi(s_t)
\end{equation}
receives the current state and returns an evaluation as the reward. 
\end{itemize}

\textbf{Reward Model.} Next, we introduce in detail the design of the reward model. The purpose of the reward model is to evaluate how the current state impacts the classification performance. Intuitively, the state which leads to better classification performance should have a higher reward: $r_t=\mathcal{F}(s_t)$. We set the reward modal $\mathcal{F}$ as a combination of the convolutional mapping and classification (Section~\ref{sub:classification}). Since in the practical approach optimization, the higher the accuracy is, the more difficult to increase the classification accuracy. For example, improving the accuracy on a higher level (e.g., from 90\% to 100\%) is much harder than on a lower level(e.g., from 50\% to 60\%). To encourage accuracy improvement at the higher level, we design a non-linear reward function:
\begin{equation}\label{eq:reward}
r_t = \frac{e^{acc}}{e-1}-\beta \frac{\bar{K}}{K'}
\end{equation}
where $acc$ denotes the classification accuracy. The function contains two parts; the first part is a normalized exponential function with the exponent $acc \in[0,1]$, this part encourages the reinforcement learning algorithm to search the better $s_t$ which leads to a higher $acc$. The motivation of the exponential function is that: \textit{the reward growth rate} is increasing with the accuracy's increase\footnote{For example, for the same accuracy increment 10\%, $acc: 90\%\rightarrow 100\%$ can earn a higher reward increment than $acc: 50\%\rightarrow 60\%$.}. The second part is a penalty factor for the attention zone length to keep the bar shorter and the $\beta$ is the penalty coefficient.

In summary, the aim of the deep reinforcement learning is to learn
the optimal attention zone $\mathbf{\bar{x}}^*_i$ which leads to the maximum reward. The selective mechanism totally iterates $N=n_e*n_s$ times where $n_e$ and $n_s$ denote the number of episodes and steps \cite{wang2015dueling}, respectively. $\varepsilon$-greedy method \cite{tokic2010adaptive} is employed in the state transition, which chooses a random action with probability $1 - \varepsilon$ or an action
according to the optimal Q function $argmax_{a_t\in\mathcal{A}}Q(s_t, a_t)$ with probability $\varepsilon$. In formula, 
\begin{equation}
a_{t+1}=\left\{\begin{matrix}
argmax_{a_t\in\mathcal{A}}Q(s_t, a_t) & \varepsilon' <\varepsilon \\ 
\bar{a}\in\mathcal{A} & otherwise
\end{matrix}\right.
\end{equation}
where $\varepsilon'\in[0,1]$ is random generated for each iteration while $\bar{a}$ is random selected in $\mathcal{A}$.

For better convergence and quicker training, the $\varepsilon$ is gradually increasing with the iterating. 
The increment $\varepsilon_0$ follows:
\begin{equation}
\varepsilon_{t+1} =\varepsilon_t +\varepsilon_0 N
\end{equation}

\textbf{Agent Policy and Optimization.} The Dueling DQN (Deep Q Networks \cite{wang2015dueling}) is employed as the optimization policy $\pi(s_t,a_t)$, which is enabled to learn the state-value function efficiently. The primary reason we employ a dueling DQN to uncover the best attention zone is that it updates all the four Q values at every step while other policies only update one Q value at each step. The Q function measures the expected sum of future rewards when taking that action and following the optimal policy thereafter. In particular, for the specific step $t$, we have:
\begin{equation}
\begin{split}
Q(s_t, a_t) &= \mathbb{E}(r_{t+1}+\gamma r_{t+2}+\gamma^2 r_{t+3}\dots) \\
&= \sum_{n=0}^{\infty}\gamma^kr_{t+k+1}
\end{split}
\end{equation}
where $\gamma\in[0,1]$ is the decay parameter that trade-off the importance of immediate and future rewards while $n$ denotes the number of following step. The value function $V(s_t)$ estimate the expected reward
when the agent is in state s. The Q function is related to the pair $(s_t, a_t)$ while the value function only associate with $s_t$.

Dueling DQN learns the Q function through the value function $V(s_t)$ and the advantage function $A(s_t,a_t)$ and combines them by the following formula 
\begin{equation}
Q(s_t, a_t) = \theta V(s_t)+ \theta' A(s_t,a_t)
\label{eq:1}
\end{equation}
where $\theta, \theta' \in \Theta$ are parameters in the dueling DQN network and are optimized automatically. Equation:~\ref{eq:1} is unidentifiable which can be observed by the fact that we can not recover $V(s_t)$ and $A(s_t,a_t)$ uniquely with the given $Q(s_t, a_t)$. To address this issue, we can force the advantage function equals to zero at the chosen action. That is, we let the network implement the forward mapping:
\begin{equation}
  Q(s_t,a_t)=V(s_t)+[A(s_t,a_t)-\max\limits_{a_{t+1}\in\mathcal{A}}(A(s_t,a_{t+1}))]
\end{equation}
Therefore, for the specific action $a*$, if 
\begin{equation}
  argmax_{a_{t+1}\in\mathcal{A}} Q(s_t,a_{t+1}) = argmax_{a_{t+1}\in\mathcal{A}} A(s_t,a_{t+1})
\end{equation}
then we have
\begin{equation}
  Q(s_{t+1}, a*)=V(s_t)
\end{equation}
Thus, as shown in the Figure~\ref{fig:workflow} (the second last layer of the agent part), the stream $V(s_t)$ is forced to learn an estimation of the value function, while the other stream produces
an estimation of the advantage function.

To assess the Q function, we optimize the following cost function at the $i$-th iteration:
\begin{equation}
\begin{split}
L_i(\Theta_i)&=\mathbb{E}_{s_t,a_t,r_t,s_{t+1}}[(\bar{y}_i-Q(s_t, a_t))^2] \\
             &=\mathbb{E}_{s_t,a_t,r_t,s_{t+1}}[(\bar{y}_i-\theta V(s_t)+ \theta' A(s_t,a_t))^2]
\end{split}
\end{equation}
with 
\begin{equation}
  \bar{y}_i=r_t + \gamma \max\limits_{a_{t+1}}Q(s_{t+1}, a_{t+1})
\end{equation}
The gradient update method is 
\begin{equation}
\begin{split}
\nabla_{\Theta_i}L_i(\Theta_i)&=\mathbb{E}_{s_t,a_t,r_t,s_{t+1}}[(\bar{y}_i-Q(s_t, a_t))\nabla_{\Theta_i}Q(s_t, a_t)] \\
&=\mathbb{E}_{s_t,a_t,r_t,s_{t+1}}[(\bar{y}_i-\theta V(s_t)- \theta' A(s_t,a_t))\\
&\nabla_{\Theta_i}(\theta V(s_t)+ \theta' A(s_t,a_t))]
\end{split}
\end{equation}

\subsection{Convolutional Mapping} 
\label{sub:classification}
For each attention zone, we further exploit the potential spatial dependency of selected features $\mathbf{\bar{x}}^*_i$. Since we focus on a single sample, the EEG sample only contains a numerical vector with very limited information and is easily corrupted by noise. To amend this drawback, we attempt to mapping the EEG single sample from the original space $\mathcal{O}\in R^K$ to a sparsity space $\mathcal{T}\in R^M$ by a CNN structure. 

To extract as more potential spatial dependencies as possible, we employ a convolutional layer \cite{krizhevsky2012imagenet} with many filters to scan on the learned attention zone $\mathbf{\bar{x}}^*_i$. The convolutional mapping structure contains five layers (as shown in Figure~\ref{fig:workflow}): the input layer receives the learned attention zone, the convolutional layer followed by one fully connected layer, and the output layer. The one-hot ground truth is compared with the output layer to calculate the training loss.

The Relu non-linear activation function is applied to the convolutional outputs. We describe the convolutional layer as follows:
\begin{equation}
  x^c_{ij} = ReLU(\sum_{b=1}^{\bar{b}}W_c \bar{x}^*_{ij})
\end{equation}
where $x^c_{ij}$ denotes the outcome of the convolutional layer while $\bar{b}$ and $W_c$ denote the length of filter and the filter weights, respectively.
The pooling layer aims to reduce the redundant information in the convolutional outputs to decrease the computational cost. In our case, we try to keep as much information as possible. Therefore, our method does not employ a pooling layer. Then, in the fully connected layer and output layer
\begin{equation}
  x^f_i = ReLU(W^f x^c_i+b^f)
\end{equation}
\begin{equation}
   y'_i = softmax(W^o x^f_i+b^o)
\end{equation}
where $W^f, W^o,b^f,b^o$ denote the corresponding weights and biases, respectively. The $y'$ denotes the predicted label.
The cost function is measured by cross entropy, and the $\ell_2$-norm (with parameter $\lambda$) is adopted as regularization to prevent overfitting.:
\begin{equation}
  cost=-\sum_{x} y'_ilog(y_i) + \lambda \ell_2
\end{equation}

The AdamOptimizer algorithm optimizes the cost function. The fully connected layer extracts as the represented features and fed them into a lightweight nearest neighbor classifier.  The convolutional mapping updates for $N'$ iterations. The proposed adaptive cognitive activity recognition with reinforced attentive convolutional neural networks is shown in Algorithm~\ref{alg:approach}.

\begin{algorithm}[!t]
\begin{small}
\caption{The Proposed Approach}
\label{alg:approach}
\begin{algorithmic}[1]
\renewcommand{\algorithmicrequire}{\textbf{Input:}}
\renewcommand{\algorithmic}{\textbf{Hyper-parameters:}}
 \renewcommand{\algorithmicensure}{\textbf{Output:}}
 \REQUIRE Raw EEG signals $\mathbf{X}$
 \ENSURE  Predicted cognitive activity label $y'_i$
 \STATE  Initialization $s_0$\;
 \STATE \textbf{RS}: $\mathbf{\bar{x}}_i\leftarrow \mathbf{x}'_i$\;
 \STATE \textbf{Reinforced Selective Attention}:
  \IF{$t<N$}
      \STATE $a_t= argmax_{a_t\in\mathcal{A}} Q(s_t,a_t)$
      \STATE $s_{t+1}=\pi(s_t,a_t)$
      \STATE $r_{t}=\mathcal{F}(s_t)$
      \STATE $\varepsilon_{t+1} =\varepsilon_t +\varepsilon_0 N$
      \STATE $\mathbf{\bar{x}}^*_i \leftarrow \mathbf{\bar{x}}_i, a_t,s_t,r_t$
  \ENDIF
  \STATE \textbf{Convolutional Mapping \& Classifier}:
    \IF{$iteration<N'$}
      \STATE $ y'_i \leftarrow \mathbf{\bar{x}}^*_i$
  \ENDIF
  \RETURN $y'_i$
\end{algorithmic}
\end{small}
\vspace{-1mm}
\end{algorithm}

\section{Experiments} 
\label{sec:experiments}
In this section, we report our evaluation of the proposed approach on three datasets corresponding to different application scenarios, with a focus on accuracy, latency, and resilience.

\begin{table*}[]
\centering
\caption{Comparison with baselines}
\label{tab:baseline}
\resizebox{0.9\textwidth}{!}{
\begin{tabular}{l|l|l|lllll|llll}
\hline
\multirow{2}{*}{\textbf{Scenarios}} & \multicolumn{1}{l|}{\multirow{2}{*}{\textbf{Datasets}}} & \multicolumn{1}{l|}{\multirow{2}{*}{\textbf{Metrics}}} & \multicolumn{5}{c|}{\textbf{Non-Deep Learning Baselines}} & \multicolumn{4}{c}{\textbf{Deep Learning Baselines}} \\ \cline{4-12}  
& \multicolumn{1}{l|}{} & \multicolumn{1}{l|}{} & \textbf{SVM} & \textbf{RF} & \textbf{KNN} & \textbf{AB} & \textbf{LDA} & \textbf{LSTM} & \textbf{GRU} & \textbf{CNN} & \textbf{Ours} \\ \hline
\multirow{4}{*}{\textbf{MIR}} & \multirow{4}{*}{\textbf{eegmmidb}} & \textbf{Accuracy} & 0.5596 & 0.6996 & 0.5814 & 0.3043 & 0.5614 & 0.648 & 0.6786 & 0.91 & \textbf{0.9632} \\
 &  & \textbf{Precision} & 0.5538 & 0.7311 & 0.6056 & 0.2897 & 0.5617 & 0.6952 & 0.8873 & 0.9104 & 0.9632 \\
 &  & \textbf{Recall} & 0.5596 & 0.6996 & 0.5814 & 0.3043 & 0.5614 & 0.6446 & 0.6127 & 0.9104 & 0.9632 \\
 &  & \textbf{F1-score} & 0.5396 & 0.6738 & 0.5813 & 0.2037 & 0.5526 & 0.6619 & 0.7128 & 0.9103 & 0.9632 \\
  \hline
\multirow{4}{*}{\textbf{PI}} & \multirow{4}{*}{\textbf{EEG-S}} & \textbf{Accuracy} & 0.6604 & 0.9619 & 0.9278 & 0.35 & 0.6681 & 0.9571 & 0.9821 & 0.998 & \textbf{0.9984} \\
 &  & \textbf{Precision} & 0.6551 & 0.9625 & 0.9336 & 0.3036 & 0.6779 & 0.9706 & 0.9858 & 0.998 & 0.9984 \\
 &  & \textbf{Recall} & 0.6604 & 0.962 & 0.9279 & 0.35 & 0.6681 & 0.9705 & 0.9857 & 0.998 & 0.9984 \\
 &  & \textbf{F1-score} & 0.6512 & 0.9621 & 0.9282 & 0.2877 & 0.668 & 0.9705 & 0.9857 & 0.998 & 0.9984 \\ \hline
\multirow{4}{*}{\textbf{ND}} & \multirow{4}{*}{\textbf{TUH}} & \textbf{Accuracy} & 0.7692 & 0.92 & 0.9192 & 0.5292 & 0.7675 & 0.6625 & 0.6625 & 0.9592 & \textbf{0.9975} \\
 &  & \textbf{Precision} & 0.7695 & 0.9206 & 0.923 & 0.7525 & 0.7675 & 0.6538 & 0.6985 & 0.9593 & 0.9975 \\
 &  & \textbf{Recall} & 0.7692 & 0.92 & 0.9192 & 0.5292 & 0.7675 & 0.6417 & 0.6583 & 0.9592 & 0.9975 \\
 &  & \textbf{F1-score} & 0.7692 & 0.9199 & 0.9188 & 0.3742 & 0.7675 & 0.6449 & 0.6685 & 0.9592 & 0.9975 \\ \hline
\end{tabular}
}
\end{table*}

\begin{table*}[]
\centering
\caption{Comparison with the state-of-the-art approaches}
\label{tab:comparison_state}
\resizebox{0.9\textwidth}{!}{
\begin{tabular}{l|l|l|llllll}
\hline 
\textbf{Scenarios} & \textbf{Datasets} & \textbf{Metrics} & \multicolumn{6}{c}{\textbf{State-of-the-art}} \\ \hline
\multirow{10}{*}{\textbf{MIR}} & \multirow{10}{*}{\textbf{eegmmidb}} & \textbf{Method} & \textbf{Rashid \cite{or2016classification}} & \textbf{Zhang \cite{zhang2017converting}} & \textbf{Ma \cite{ma2018improving}} & \textbf{Alomari \cite{Alomari2014}} & \textbf{Sita \cite{sita2013feature}} & \textbf{Alomari \cite{alomari2014eeg}} \\
 &  & \textbf{Accuracy} & 0.9193 & 0.9561 & 0.6820 & 0.8679 & 0.7584 & 0.8515 \\
 &  & \textbf{Precision} & 0.9156 & 0.9566 & 0.6971 & 0.8788 & 0.7631 & 0.8469 \\
 &  & \textbf{Recall} & 0.9231 & 0.9621 & 0.7325 & 0.8786 & 0.7702 & 0.8827 \\
 &  & \textbf{F1-score} & 0.9193 & 0.9593 & 0.7144 & 0.8787 & 0.7666 & 0.8644 \\ \cline{3-9}
 &  & \textbf{Method} & \textbf{Shenoy \cite{shenoy2015shrinkage}} & \textbf{Szczuko \cite{szczuko2017real}} & \textbf{Stefano\cite{stefano2017eeg}} & \textbf{Pinheiro \cite{pinheiro2016wheelchair}} & \textbf{Kim \cite{kim2016motor}} & \textbf{Ours} \\
 &  & \textbf{Accuracy} & 0.8308 & 0.9301 & 0.8724 & 0.8488 & 0.8115 & \textbf{0.9632} \\
 &  & \textbf{Precision} & 0.8301 & 0.9314 & 0.8874 & 0.8513 & 0.8128 & 0.9632 \\
 &  & \textbf{Recall} & 0.8425 & 0.9287 & 0.8874 & 0.8569 & 0.8087 & 0.9632 \\
 &  & \textbf{F1-score} & 0.8363 & 0.9300 & 0.8874 & 0.8541 & 0.8107 & 0.9632 \\ \hline 
\multirow{5}{*}{\textbf{PI}} & \multirow{5}{*}{\textbf{EEG-S}} & \textbf{Method} & \textbf{Ma \cite{ma2015resting}} & \textbf{Yang \cite{yang2014novel}} & \textbf{Rodrigues \cite{rodrigues2016eeg}} & \textbf{Frashini \cite{fraschini2015eeg}} & \textbf{Thomas \cite{thomas2016biometric}} & \textbf{Ours} \\
 &  & \textbf{Accuracy} & 0.88 & 0.99 & 0.8639 & 0.956 & 0.9807 & \textbf{0.9984} \\
 &  & \textbf{Precision} & 0.8891 & 0.9637 & 0.8721 & 0.9458 & 0.9799 & 0.9984 \\
 &  & \textbf{Recall} & 0.8891 & 0.9594 & 0.8876 & 0.9539 & 0.9887 & 0.9984 \\
 &  & \textbf{F1-score} & 0.8891 & 0.9615 & 0.8798 & 0.9498 & 0.9843 & 0.9984 \\ \hline
\multirow{5}{*}{\textbf{ND}} & \multirow{5}{*}{\textbf{TUH}} & \textbf{Method} & \textbf{Ziyabari \cite{ziyabari2017objective}} & \textbf{Harati \cite{harati2015improved}} & \textbf{Zhang \cite{zhang2018integration} } & \textbf{Goodwin \cite{goodwin2017deep}} & \textbf{Golmohammadi \cite{golmohammadi2017automatic}} & \textbf{Ours} \\
 &  & \textbf{Accur{}acy} & 0.9382 & 0.9429 & 0.994 & 0.924 & 0.9479 & \textbf{0.9975} \\
 &  & \textbf{Precision} & 0.9321 & 0.9503 & 0.9951 & 0.9177 & 0.9438 & 0.9975 \\
 &  & \textbf{Recall} & 0.9455 & 0.9761 & 0.9951 & 0.9375 & 0.9522 & 0.9975 \\
 &  & \textbf{F1-score} & 0.9388 & 0.9630 & 0.9951 & 0.9275 & 0.9480 & 0.9975 \\ \hline
\end{tabular}
}
\vspace{-3mm}
\end{table*}

\subsection{Application Scenarios and Datasets} 
\label{sec:application_scenarios_and_datasets}

\subsubsection{Application Scenarios}
We evaluate our approach on various datasets in three applications of EEG-based Brain-Computer Interfaces.

\vspace{0.1cm}
\noindent\textbf{Movement Intention Recognition (MIR).} EEG signals measure human brain activities. Intuitively, different human intention will lead to diverse EEG patterns \cite{zhang2017converting}. Intention recognition plays a significant role in practical scenarios such as smart home, assisted living \cite{zhang2017intent}, brain typing \cite{zhang2017converting}, and entertainment. For the disabled and elders, intent recognition can help them to interact with external smart devices such as wheelchairs or service robots real-time BCI systems. Besides, for people without vocal ability, they may have the chance to express their thoughts with the help of certain intention recognition technologies (e.g., brain typing). Even for the healthy human being, intent recognition can be used in video game playing and other daily living applications. 

\vspace{0.1cm}
\noindent\textbf{Person Identification (PI).} EEG-based biometric identification \cite{schetinin2017feature} is an emerging person identification approach, 
which is highly attack-resilient. It has the unique advantage of avoiding or alleviating the threat of being deceived which is often faced by other identification techniques. This technique can be deployed in identification and authentication scenarios such as bank security system and customs security check.

\vspace{0.1cm}
\noindent\textbf{Neurological Diagnosis (ND).} EEG signals collected in the unhealthy state differ significantly from the ones collected in the normal state concerning frequency and pattern of neuronal firing \cite{adeli2007wavelet}. Therefore, EEG signals have been used for neurological diagnosis for decades \cite{zhang2019adversarial}. For example, the epileptic seizure is a common brain disorder that affects around 1\% of the population, and an EEG analysis of the patient could detect its octal state.

\subsubsection{Datasets}
To evaluate how the proposed approach works in the aforementioned application scenarios, we choose several EEG datasets with various collection equipment, sampling rates, and data sources.
We utilize motor imagery EEG signals from a public dataset \textit{eegmmidb} for intention recognition,
the \textit{EEG-S} dataset for person identification, and the \textit{TUH} dataset for neurological diagnosis. 

\vspace{0.1cm}
\noindent\textbf{eegmmidb.} EEG motor movement/imagery database (eegmmidb)\footnote{\url{https://www.physionet.org/pn4/eegmmidb/}} were collected by the BCI200 EEG system, which recordsed the brain signals using 64 channels with a sampling rate of 160Hz. EEG signals were recorded when the subject was imaging about certain actions (without any physical action). This dataset includes 560,000 samples collected from 20 subjects. Each sample have one of five different labels: eye-closed, left hand, right hand, both hands, and both feed. Each sample is a vector of 64 elements that correspond to 64 channel of EEG data. 

\vspace{0.1cm}
\noindent\textbf{EEG-S.} EEG-S is a subset of eegmmidb,
 in which the data were gathered while the subject kept eyes closed and stayed relaxed. Eight subjects were involved and each subject generated 7,000 samples. Labels are the subjects' IDs, which range within [0-7].

\vspace{0.1cm}
\noindent\textbf{TUH.} TUH \cite{golmohammadi2017tuh} is a neurological seizure dataset of clinical EEG recordings. The EEG recording is associated with 22 channels from a 10/20 configuration and a sampling rate of 250 Hz. We selected 12,000 samples from each of five subjects (2 males and three females). Half of the samples were labeled as epileptic seizure state. The remaining samples were labeled as the normal state.

\subsubsection{Parameter Settings}
We configured the default settings of our approach as follows. In the selective attention learning: $\bar{K}=128$, the Dueling DQN had 4 lays and the node number in each layer were: 2 (input layer), 32 (FCL), 4 ($A(s_t,a_t)$) + 1 ($V(s_t)$), and 4 (output). The decay parameter $\gamma =0.8$, $n_e=n_s=50$, $N=2,500$, $\epsilon=0.2$, $\epsilon_0=0.002$, learning rate$ =0.01$, memory size $ =2000$, length penalty coefficient $\beta=0.1$, and the minimum length of attention zone was set as 10. 
In the convolutional mapping, the node number in the input layer equaled to the number of attention zone dimensions. In the convolutional layer: the stride had the shape $[1,1]$, the filter size was set to $[1,2]$, the depth to 10, and the non-linear function as ReLU. The padding method was zero-padding. No pooling layer was adopted. The subsequent fully connected layer had 100 nodes. The learning rate was $0.001$ while the $\ell_2$-norm coefficient $\lambda$ equaled 0.001. The transformation was trained for 2000 iterations. 
In addition, we configured the key parameters of the baselines as follows: Linear SVM ($C=1$), Random Forest (RF, $n=200$), KNN ($k=1$). In LSTM (Long Short-Term Memory) and GRU (Gated Recurrent Unit), $n_{steps}=5$, other settings were the same as \cite{zhang2017intent}. The CNN had the same structure and hyper-parameters setting with the convolutional mapping component in the proposed show.


\begin{figure*}[]
    \centering
    \hspace{3mm}
    \begin{subfigure}[t]{0.28\textwidth}
        \centering
        \includegraphics[width=\textwidth]{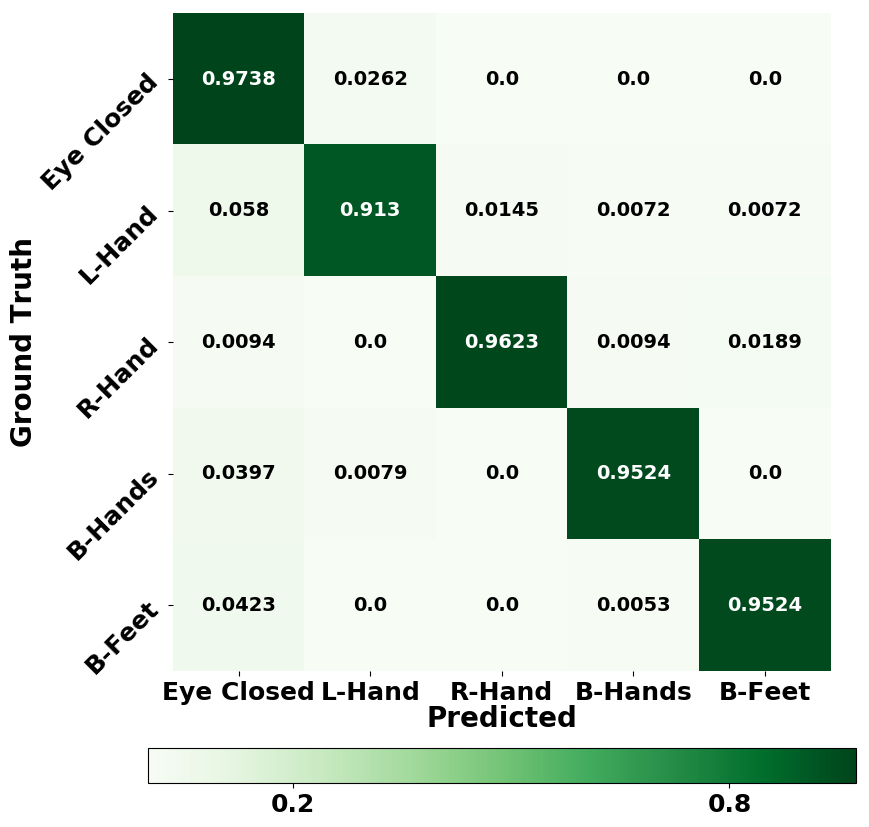}
        \caption{CM of eegmmidb}
        \label{fig:cm_eegmmidb}
    \end{subfigure}%
  \hspace{5mm}
    \begin{subfigure}[t]{0.25\textwidth}
        \centering
        \includegraphics[width=\textwidth]{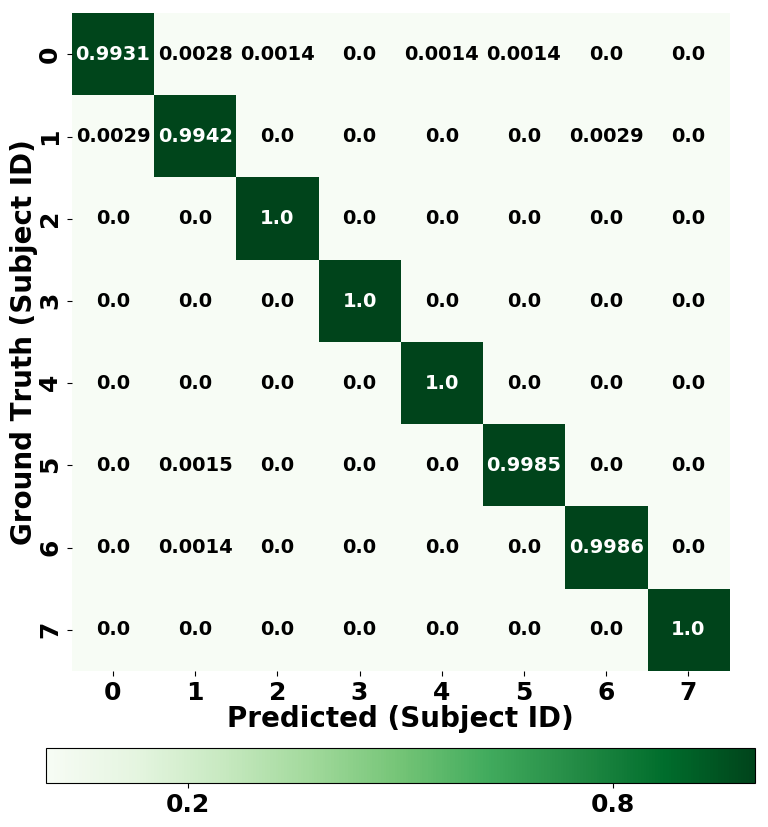}
        \caption{CM of EEG-S}
    \end{subfigure}%
    \hspace{5mm}
    \begin{subfigure}[t]{0.27\textwidth}
        \centering
        \includegraphics[width=\textwidth]{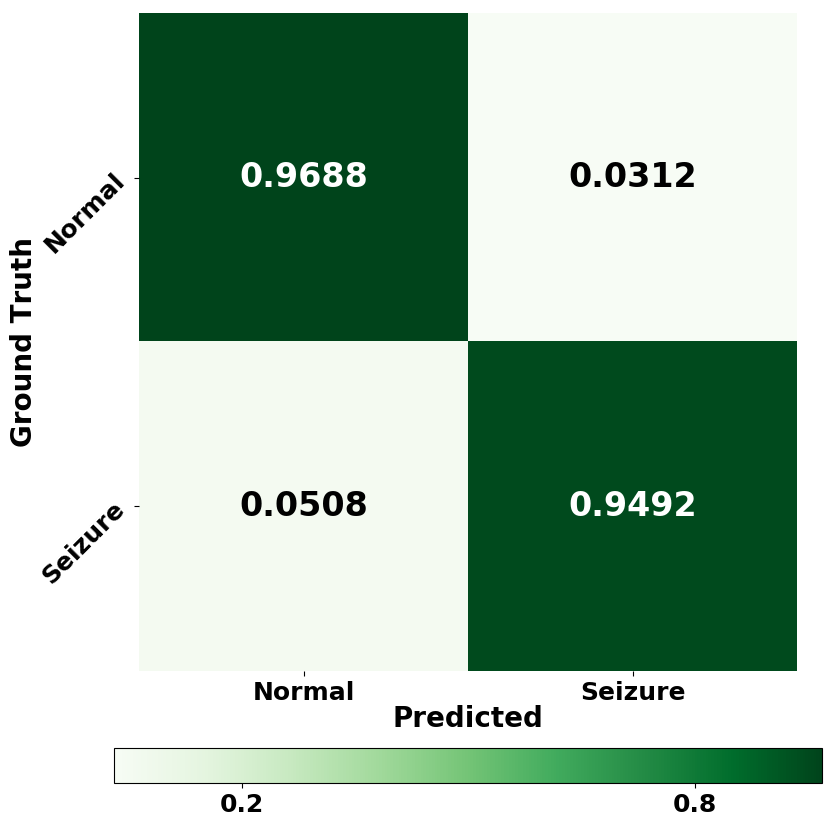}
        \caption{CM of TUH}
    \end{subfigure}%

    \centering
    \hspace{4mm}
    \begin{subfigure}[t]{0.25\textwidth}
        \centering
        \includegraphics[width=\textwidth]{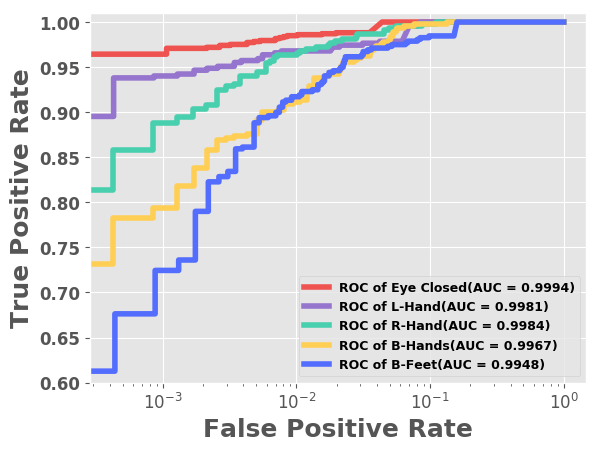}
        \caption{ROC of eegmmidb}
    \end{subfigure}%
    \hspace{8mm}
    \begin{subfigure}[t]{0.25\textwidth}
        \centering
        \includegraphics[width=\textwidth]{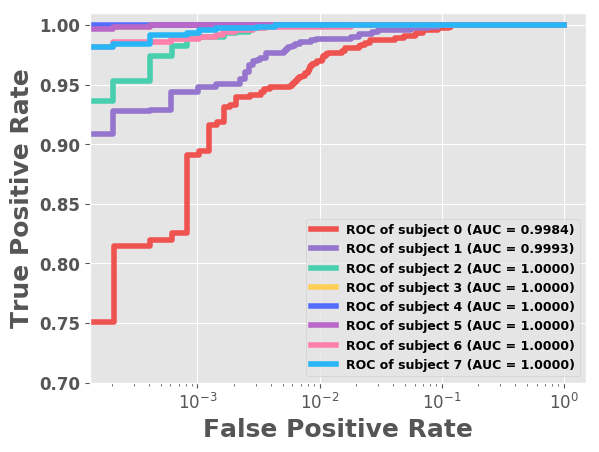}
        \caption{ROC of EEG-S}
    \end{subfigure}%
    \hspace{10mm}
    \begin{subfigure}[t]{0.25\textwidth}
        \centering
        \includegraphics[width=\textwidth]{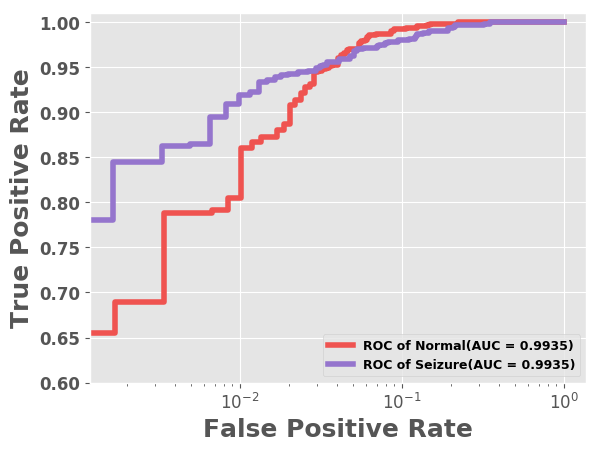}
        \caption{ROC of TUH}
    \end{subfigure}%
    \caption{Confusion matrix and ROC curves with AUC scores of each dataset. CM denotes confusion matrix. 
    }
    \label{fig:cm_roc}
    \vspace{-3mm}
\end{figure*}

\subsection{Overall Comparison} 
\label{sec:approach_comparison}

\subsubsection{Comparison Baselines} 
\label{sub:comparison_baselines}

To measure the accuracy of the proposed method, we compared with a set of baseline methods including five non-deep learning and three deep learning based baselines. Furthermore, we chose some competitive state-of-the-art algorithms for every single task separately.

\vspace{2mm}
\noindent\textbf{MIR Baselines:} 

Rashid et al. \cite{or2016classification} use Discrete Wavelet Transform (DWT) to extract features and feed into Levenberg-Marquardt Algorithm (LMA) based neural network for motor imagery EEG intention recognition.

Zhang et al. \cite{zhang2017converting} design a joint convolutional recurrent neural network to learn robust high-level feature presentations by low-dimensional dense embeddings from raw MI-EEG signals.


Ma et al. \cite{ma2018improving} transform the EEG data into a spatial sequence to learn more valuable information through RNN. 

Alomari et al. \cite{Alomari2014} analyze the EEG characteristics by the Coiflets wavelets and manually extract features using different amplitude estimators. The extracted features are inputted into SVM classifier for EEG data recognition.

Sita et al. \cite{sita2013feature} employ independent component analysis (ICA) to extract features which are fed to a quadratic discriminant analysis (QDA) classifier.

Alomari et al. \cite{alomari2014eeg} use wavelet transformation to filter and process EEG signals. Then calculate the Root Mean Square and Mean Absolute Value features for EEG recognition. 

Shenoy et al. \cite{shenoy2015shrinkage} propose a regularization approach based on shrinkage estimation to handle small sample problem and retain subject-specific discriminative features.

Szczuko \cite{szczuko2017real} design a rough set based classifier for the aim of EEG data classification.

Stefano et al. \cite{stefano2017eeg} extract the mu ($7\sim13 Hz$) and beta ($13 \sim30 Hz$) bands' power spectral density (PSD) as manual features to discriminate different motor imagery intentions.

Pinheiro et al. \cite{pinheiro2016wheelchair} adopt a C4.5 decision tree as the classifier to distinguish the manually extracted EEG features such as arithmetic mean and maximum value of the Fourier transform.

Kim et al. \cite{kim2016motor} use a multivariate empirical mode decomposition to obtain the mu and beta rhythms from the nonlinear EEG signals.

\vspace{2mm}
\noindent\textbf{PI Baselines:} 
 
Ma et al. \cite{ma2015resting} adopt a CNN structure to automatically extract an individual's best, unique neural features with the aim of person identification.

Yang et al. \cite{yang2014novel} present an approach for biometric identification using EEG signals based on features extracted with the Hilbert-Huang Transform (HHT). 

Rodrigues et al. \cite{rodrigues2016eeg} propose the Flower Pollination Algorithm under different transfer functions to select the best subset of channels that maximizes the accuracy, which is measured using the Optimum-Path Forest classifier.

Frashini et al. \cite{fraschini2015eeg} decompose EEG signals into standard frequency bands by a band-pass filter and estimate the functional connectivity between the sensors using the Phase Lag Index. The resulting connectivity matrix was used to construct a weighted network for person identification.

Thomas et al. \cite{thomas2016biometric} extract sample entropy features from the delta, theta, alpha, beta and gamma bands of 64 channel EEG data, which are evaluated for subject-identification.
 
\vspace{2mm}
\noindent\textbf{ND Baselines:} 

Ziyabari et al. \cite{ziyabari2017objective} adopt a hybrid deep learning architecture, including LSTM and stacked denoising Autoencoder, that integrates temporal and spatial context to detect the seizure.

Harati et al. \cite{harati2015improved} demonstrate that a variant of the filter bank-based approach
and provides a substantial reduction in the overall error rate.

Zhang et al. \cite{zhang2018integration} extract a list of 24 feature types from the scalp EEG signals and found 170 out of the 2794 features to classify epileptic seizures accurately. 


Goodwin et al. \cite{goodwin2017deep} combine recent advances in RNN with access to textual data in EEG reports to automatically extracting word- and report-level features and infer underspecified information from EHRs (electronic health records).

Golmohammadi et al. \cite{golmohammadi2017automatic} propose a seizure detection method by using hidden Markov models (HMM) for sequential decoding and deep learning networks. 

\subsubsection{Results} 
\label{sub:results}
Tables~\ref{tab:baseline} presents the classification metrics comparison between our approach and well-known baselines (including Non-DL and DL baselines), where DL, AdaB, LDA represent deep learning, Adaptive Boosting, and Linear Discriminant Analysis, respectively. The results show that our approach achieved the highest accuracy on all the datasets. Specifically, the proposed approach achieved the highest accuracy of 0.9632, 0.9984, and 0.9975 on eegmmidb, EEG-S, and TUH dataset, respectively. Further, we conducted an ablation study by comparing our method, which mainly combined selective attention mechanism and CNN, with the solo CNN. It turned out that our approach outperformed CNN, demonstrating the proposed selective attention mechanism improved the distinctive feature learning.We show the confusion matrix and ROC curves (including the AUC scores) of each dataset in Figure~\ref{fig:cm_roc}. In Figure~\ref{fig:cm_eegmmidb}, `L', `R', and `B' denote left, right, and both, respectively.

Besides, to further evaluate the performance of our model, we compared our framework with 21 state-of-the-art methods which using the same dataset. In particular, we compared with 11 competitive state-of-the-art methods over motor imagery classification and five cutting edges separately over person identification and neurological diagnosis scenarios. Table~\ref{tab:comparison_state} shows the comparison results.

We could observed that our proposed framework consistently outperformed a set of widely used baseline methods and strong competitors on three different datasets. The performance shows a significant improvement compared with other baselines. These datasets were collected using different EEG hardware, ranging from high-precision medical equipment to off-the-shelf EEG headset with a different number of EEG channels.
Regarding the seizure diagnosis in ND, by setting the normal state as impostor while the seizure state as genuine, our approach gained a False Acceptance Rate (FAR) of 0.0033 and a False Rejective Rate (FRR) of 0.0017. This outperformed the existing methods by a large margin \cite{acar2007seizure,goodwin2017deep,golmohammadi2017automatic,harati2015improved}.

\begin{figure}[t]
\centering
\begin{minipage}[b]{0.49\linewidth}
\centering
\includegraphics[width=\textwidth]{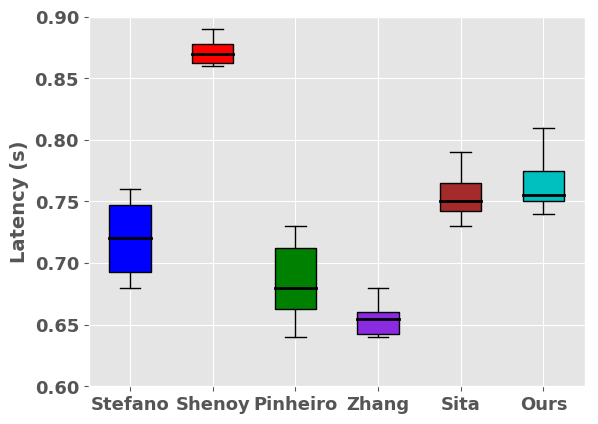}
\caption{Latency comparison}
\label{fig:latency}
\end{minipage}
\begin{minipage}[b]{0.49\linewidth}
\centering
\includegraphics[width=0.9\textwidth]{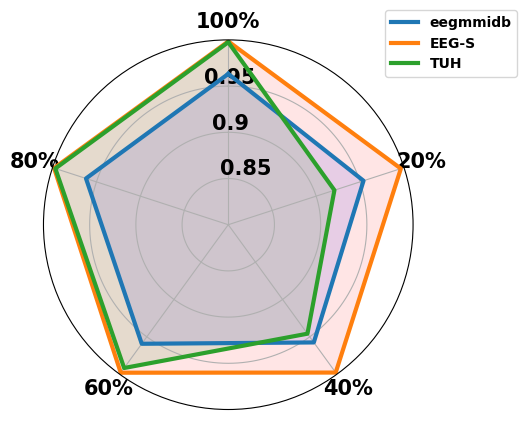}
\caption{Varying \# of channels}
\label{fig:feature}
\end{minipage}
\vspace{-5mm}
\end{figure}

\subsection{Resilience Evaluation} 
\label{sec:impact_of_feature_robustness}
In this section, we focus on evaluating the resilience of proposed method in coping with various number of EEG signal channels, and incomplete EEG signals. 

In practice, the number of EEG channels of EEG devices are diverse due to two reasons. First, different off-the-shelf or on-the-shelf devices have various channels numbers. Intuitively, the quality of signals and the contained information is directly associated with the number of channels. In the meantime, the devices with more channels usually are more expensive and less portable. 
Second, incomplete EEG signals cause the degradation of BCI applications. It could happen when some electrical nodes are loosened because of weak maintenance of EEG devices. To investigate the robustness of incomplete EEG signals with missing channels, we also conduct experiments by randomly selecting part of a proportion of signal channels over three datasets. For example, by selecting 20\% of channels on the eegmmidb dataset, the selected channel number is $12=round(64*0.2)$. Figure~\ref{fig:feature} shows the experiments results (0.4 denotes the accuracy and 20\% denotes the channel percentage used for training). The radar chart demonstrates that eegmmidb and EEG-S, both with 64 channels, can achieve competitive accuracy even with only 20\% signal channels. In contrast, TUH (22 channels) is highly dependent on the channel numbers. The reason is that TUH only remains five channels for 20\% channel percentage, respectively. According to our experience, the proposed framework requires at least eight EEG channels to achieve high accuracy.



\subsection{Latency Analysis} 
\label{sec:latency}
Except for the high accuracy of EEG signal classification, the low latency is another critical requirement for the success of real-world BCI applications. 

In this section, we take the eegmmidb dataset as an example to compare the latency of the proposed framework with several state-of-the-art algorithms.
The results are presented in Figure~\ref{fig:latency}. We observed that our approach had competitive latency compared with other methods. The overall latency was less than 1 second. The deep learning based techniques in this work do not explicitly lead to extra latency. One of the main reasons may lie in that the reinforced selective attention has filtered out unnecessary information. To be more specific, the classification latency of the proposed framework was about 0.7$\sim$0.8 seconds, which mainly resulted from the classifying procedure and convolutional mapping. The latency caused by the classifier was around 0.7 seconds. The convolutional mapping only took 0.05 sec on testing although it took more than ten minutes on training. 

\subsection{Reward Model Demonstration} 
\label{sub:reward_model_demonstration}
We briefly report the empirical demonstration of the proposed exponential reward model (Section~\ref{sub:attention_pattern_learning}). We compared the proposed reward model in Eq.~\ref{eq:reward} with the traditional reward $r_t = e^{acc}$ over three benchmark datasets (eegmmidb, EEG-S, and TUH). The experiment results show that the novel reward model achieved higher accuracy (0.9632, 0.9984, and 0.9975) than the traditional model (0.9231, 0.9901, and 0.9762).

\section{Discussions} 
\label{sub:discussion}
In this paper, we propose a robust, universal, adaptive classification framework to deal with cognitive EEG signals effectively and efficiently. 
However, there are several remaining challenges. 

First, the single EEG sample-based classification can only reflect the instantaneous intention of the subject instead of a long-term stable intention. One possible modification method is post-processing like voting. For instance, we can classify 100 EEG samples and count the mode of the 100 outcomes as the final classification result. 

In addition, the reinforcement learning policy only works well in the environment in which the model is trained, meaning the dimension index should be consistent in the training and testing stages. Various policies should be trained according to different sensor combinations. Also, the replicate and shuffle process cannot always provide the best spatial dependency. Therefore, when the classification accuracy is not satisfied, repeating the replicate and shuffle procedure help to enhance the additional performance.



\section{Conclusion} 
\label{sec:discussions_and_conclusion}
This paper proposes a generic and effective framework for raw EEG signal classification to support the development of BCI applications. The framework works directly on raw EEG data without requiring any preprocessing or feature engineering. Besides, it can automatically select distinguishable feature dimensions for different EEG data, thus achieving high usability. We conduct extensive experiments on three well-known public datasets and one local dataset. The experimental results demonstrate that our approach not only outperforms several state-of-the-art baselines by a large margin but also shows low latency and high resilience in coping with multiple EEG signal channels and incomplete EEG signals. Our approach applies to wider application scenarios such as intention recognition, person identification, and neurological diagnosis.

\section{ACKNOWLEDGMENTS} 
This research was partially supported by grant ONRG NICOP N62909-19-1-2009.

\bibliographystyle{IEEEtran}
\bibliography{icdm2019}

\end{document}